\documentclass[aps,twocolumn,prb,showpacs,superscriptaddress,floatfix]{revtex4}
\usepackage{graphicx,amsmath}
\usepackage{wasysym,pifont}
\usepackage{epsfig}
\usepackage{newlfont,bm,dcolumn,longtable}
\usepackage{amssymb,amsfonts,amsmath}
\usepackage{color}

\begin{document}

\newcommand{\be}{\begin{equation}}
\newcommand{\ee}{\end{equation}}
\newcommand{\bea}{\begin{eqnarray}}
\newcommand{\eea}{\end{eqnarray}}
\newcommand{\<}{\langle}
\renewcommand{\>}{\rangle}
\newcommand{\qea}{q_{\scriptscriptstyle{\rm EA}}}
\newcommand{\teff}{t_{\rm eff}}
\newcommand{\thav}[1]{\langle #1 \rangle}
\newcommand{\smav}[1]{\overline{#1}}
\newcommand{\average}[1]{\smav{\thav{#1}}}
\title{
Weak first-order transition in the three-dimensional site-diluted Ising
antiferromagnet in a magnetic field 
}

\author{A.~Maiorano}
\affiliation{Departamento de F\'{\i}sica, Facultad de Ciencias,\\
Universidad de Extremadura, 06071 Badajoz, SPAIN}
\affiliation{Instituto de Biocomputaci\'on y  F\'{\i}sica de
Sistemas Complejos (BIFI),\\
Facultad de Ciencias, Universidad de Zaragoza, 50009 Zaragoza, SPAIN}
\author{V.~Mart\'{\i}n-Mayor}
\affiliation{Departamento de F\'{\i}sica Te\'orica, Facultad de
  Ciencias F\'{\i}sicas,\\
Universidad Complutense de Madrid, 28040 Madrid, SPAIN}
\affiliation{Instituto de Biocomputaci\'on y  F\'{\i}sica de
Sistemas Complejos (BIFI),\\
Facultad de Ciencias, Universidad de Zaragoza, 50009 Zaragoza, SPAIN}
\author{J.~J.~Ruiz-Lorenzo}
\affiliation{Departamento de F\'{\i}sica, Facultad de Ciencias,\\
Universidad de Extremadura, 06071 Badajoz, SPAIN}
\affiliation{Instituto de Biocomputaci\'on y  F\'{\i}sica de
Sistemas Complejos (BIFI),\\
Facultad de Ciencias, Universidad de Zaragoza, 50009 Zaragoza, SPAIN}
\author{A.~Taranc\'on}
\affiliation{Departamento de F\'{\i}sica Te\'orica,
Facultad de Ciencias, \\
Universidad de Zaragoza, 50009 Zaragoza, SPAIN}
\affiliation{Instituto de Biocomputaci\'on y  F\'{\i}sica de
Sistemas Complejos (BIFI),\\
Facultad de Ciencias, Universidad de Zaragoza, 50009 Zaragoza, SPAIN}
\date{\today}

\date{\today}

\begin{abstract}
  
  We perform intensive numerical simulations of the three-dimensional
  site-diluted Ising antiferromagnet in a magnetic field at high values of the
  external applied field. Even if data for small lattice sizes are compatible
  with second-order criticality, the critical behavior of the system shows a
  crossover from second-order to first-order behavior for large system sizes,
  where signals of latent heat appear. We propose ``apparent'' critical
  exponents for the dependence of some observables with the lattice size for a
  generic (disordered) first-order phase transition.

\end{abstract}


\maketitle

\section{INTRODUCTION}

The study of systems with random fields is of paramount importance in
the arena of the disordered systems. A paradigm in this field is the
random field Ising model~\cite{belanger,nattermann} (RFIM).  In spite
of much effort devoted to the investigation of the
RFIM,~\cite{nattermann} several important questions remain open.  Some
of these questions refer to the nature of the (replica symmetric?) low
temperature phase, to universality issues (binary versus Gaussian
external magnetic field~\cite{sourlas,parisi-sourlas}), and to the order of the phase transitions.
Here we study the diluted antiferromagnetic Ising model in an external
magnetic field (DAFF) that is believed to belong to the same
universality class of the RFIM (the DAFF is expected to behave as a
{\em Gaussian} RFIM because of the short ranged correlations in the
superexchange coupling).~\cite{aharony_II,cardy}

As a matter of fact, DAFF systems are the most widely investigated
experimental realization of the RFIM.  One of the best examples of a diluted
Ising antiferromagnet is $\mathrm{Fe}_x \mathrm{Zn}_{1-x} \mathrm{F}_2$. Its
large crystal field anisotropy persists even when the $\mathrm{Fe}$ ions are
diluted ($x < 1$), thus providing a good (antiferromagnetic) Ising system for
all ranges of the magnetic concentration. Other systems behaving as Ising
(anti)ferromagnets are $\mathrm{Fe}_x \mathrm{Mg}_{1-x} \mathrm{Cl}_{2}$,
$\mathrm{Co} \mathrm{Zn}_{1-x} \mathrm{F}_2$ and $\mathrm{Mn}_x
\mathrm{Zn}_{1-x} \mathrm{F}_2$.~\cite{belanger} The experimental results on
the order of the phase transition are somewhat inconclusive. On the one hand,
these materials show a large critical slowing down around the critical
temperature, as well as a symmetric logarithmic divergence of the specific
heat. On the other hand, the order parameter (the staggered magnetization)
behaves with a critical exponent $\beta$ near zero, possibly marking the onset
of a first-order phase transition.~\cite{belanger} Note that $\beta$ should be
{\em exactly} zero if the order parameter is discontinuous as a function of
temperature or the magnetic field.

The numerical investigations of the DAFF at $T > 0$ are scarce. It was
investigated a long time ago by Ogielski and Huse.~\cite{ogielski} They considered
several values of the pair temperature--magnetic field on lattice
sizes up to $L=32$ but far from the critical region.~\cite{fn1} They investigated the
thermodynamics as well as the (equilibrium) dynamical critical
behavior. Their thermodynamic results pointed to a second-order phase
transition. However, they found activated dynamics (which could be
interpreted as a signal of a first-order phase transition) rather than
standard critical slowing down (for numerical studies of the DAFF at
$T=0$, see Refs. [\onlinecite{sourlas}] and [\onlinecite{hartmann1}]). 

Numerical and analytical studies rather focused on the RFIM, which is
expected to display the DAFF critical
behavior.~\cite{aharony_II,cardy} Even if the RFIM is more amenable
than the DAFF to analytical investigations, the situation is still
confusing. Indeed, mean-field theory predicts a second-order phase
transition for low magnetic field. If the probability distribution
function of the random fields does {\em not} have a minimum at zero
field, the transition is expected to remain of the second-order all
the way down to zero temperature. However, if the
probability distribution for the random field does show a minimum at
zero field, a tricritical point and a first-order 
transition line at sufficiently high field values are
predicted.~\cite{aharony_I}

The numerical investigation of the RFIM has neither confirmed nor
refuted this counterintuitive mean-field result. Rieger and Young
studied binary distributed quenched magnetic fields,~\cite{rieger_II}
where mean-field predicts a tricritical point.  After extrapolation to
the thermodynamic limit, they interpreted their results as
indicative of a first-order transition for \emph{all} external field
strengths in the thermodynamic limit (the tricritical point did not
show up).  Rieger studied the case with Gaussian
fields,~\cite{rieger_I} where only second-order behavior should be
found according to mean-field expectations. Actually, his results
were consistent with a second-order phase transition with vanishing
(!)  order parameter exponent. The simulation of Hernandez and
Diep~\cite{hernandez} of the binary RFIM supports the existence of a
tricritical point at finite temperature and magnetic field.  Also the
study by Machta \emph{et al.}~\cite{Machta} of the Gaussian RFIM
showed evidences of finite jumps in the magnetization at disorder
dependent transition points.

A completely different numerical strategy is suggested by the
expectations of a $T\!=\!0$ renormalization-group
fixed point~\cite{BRAYMOORE}. Since the ground state for a RFIM on a
sample of linear size $L$ can be found in polynomial (in $L$) time,
$T\!=\!0$ physics can be directly addressed by studying the properties of
the ground state for a large number of samples. Hartmann and
Young~\cite{hartmann2} studied lattices up to $L\!=\!96$ for the gaussian
RFIM.  They concluded that their data supported a second-order phase
transition scenario for the Gaussian RFIM. The same model was
investigated by Middleton and Fisher~\cite{middleton} in lattices up
to $L\!=\!256$. Their data suggested as well a continous phase transition
with a {\em very} small order parameter exponent, $\beta\!=\!0.017(5)$.
Using the same technique Hartmann and Nowak~\cite{hartmann1} found
non-universal behavior in the binary and Gaussian RFIM, not excluding
that the former could undergo a first-order phase transition. They
also studied $T\!=\!0$ critical properties of the DAFF for system sizes up
to $L\!=\!120$, and found critical exponents $\beta\!=\!0.02(1)$,
$\nu\!=\!1.14(10)$ and $\overline{\gamma}\!=\!3.4(4)$, compatible with their
results for the Gaussian RFIM.

The aim of this work is to revisit the Ogielski-Huse
investigation, that was carried out for $T>0$, with modern computers,
algorithms~\cite{pt} and finite-size scaling methods (the quotient
method,~\cite{quotient} that uses the finite volume correlation
length~\cite{cooper} to characterize the phase transition). The
significant CPU investment allowed us to simulate large lattices
($L\!=\!24$) and a large number of disorder realizations.  We plan, in the
future, to perform a more complete investigation of the critical
surface of this model (this would require to vary three variables:
temperature, dilution and magnetic field).

Our main finding is that the DAFF probably undergoes a very weak
first-order phase transition. This seems a natural explanation for the
finding of the activated dynamics (at equilibrium) in
Ref. [\onlinecite{ogielski}].  The discontinuity in the magnetization
density is sizable, yet that of the internal energy is {\em very}
small. Nevertheless, even if we perform a standard second-order
analysis, the critical exponent for the staggered magnetization turns
out to be ridiculously small.

The outline of the rest of this paper is the following: in the next section we
describe the model (Sect.~\ref{MODEL-PHASE-DIAGRAM}), observables
(Sect.~\ref{OBSERVABLES}), as well as the theoretical expectations for the
finite-size scaling behavior in a second-order phase transition
(Sect.~\ref{FSS-2-SUBSECT}) and for a first-order one (Sect.~\ref{FSS-1-SUBSECT}).
Details about our simulations are given in Sect.~\ref{SIMULATION}. Our
numerical results are presented in Sect.~\ref{NUMERICAL}. We first perform a
second-order analysis (Sect.~\ref{NUMERICAL-1}), then consider the possibility of a
weak first-order transition (Sect.~\ref{NUMERICAL-2}).  After summarizing our
results in Sect.~\ref{CONCLUSIONS}, we discuss in the Appendix that
the bound~\cite{chayes} $\nu\ge 2/d$ of Chayes {\em et al.} holds as well for
first-order phase transitions in the presence of disorder. In addition, we have
found upper bounds for the divergence of the susceptibility and specific heat
with the lattice size.

\section{Model}\label{MODEL}

\subsection{Model, phase diagram, symmetries}\label{MODEL-PHASE-DIAGRAM}

The model is defined in terms of Ising spin variables $S_i=\pm 1$,
$i=1,\dots,V=L^3$ placed on the nodes of a cubic lattice of size $L$ with
periodic boundary condition. The spins interact through the following
lattice Hamiltonian
\be
\label{eq:DAFF_H}
\mathcal{H}=\sum_{<i,j>}\epsilon_iS_i\epsilon_jS_j-H\sum_{i}\epsilon_iS_i\,,
\ee where the first sum runs over pairs of nearest-neighbor sites, while $H$
is the external uniform magnetic field. The $\epsilon_i$ are quenched dilution
variables, taking values $0$ and $1$ (with probability $1-p$ and $p$
respectively) in {\em empty} and {\em occupied sites}. In this work we have
fixed this probability to $p=0.7$. In this way we are guaranteed to stay away
from both from the pure case ($p=1$) and the percolation threshold
($p_c\approx0.31$).~\cite{stauffer}

It is understood that for every choice of the $\{\epsilon_i\}_{i=1}^V$, called
hereafter a {\em sample} or a {\em disorder realization}, we are to perform
the Boltzmann (thermal) average.  The mean over the disorder is only taken
afterwards.

For low magnetic field, at low temperatures, model (\ref{eq:DAFF_H}) stays in
an antiferromagnetic state that we will call the ordered phase. The
staggered magnetization $M_\mathrm{s}$, see Eq.(\ref{eg:stag_mag}) below, is
an order parameter for this phase. Note that, for $H=0$, the $Z_2$
transformation $S_i\rightarrow -S_i$, yields a degenerate antiferromagnetic
state. The increase of the magnetic field or the temperature $T$ weakens the
antiferromagnetic correlations, and the system eventually enters into a
paramagnetic state. The paramagnetic and the ordered phases are separated by a
critical line in the ($T$,$H$) plane [it would be a critical surface in the
($T$,$H$,$p$) phase diagram].

Note that the effect of disorder (quenched dilution), combined with the
applied field $H$ in a finite DAFF system, breaks the $Z_2$ symmetry even in
the ordered phase. Consider the state of minimal energy at $T=0$ and $H$ low
enough so that the staggered magnetization is maximal and $M_\mathrm{s}=p$.
Now let us change the sign to all spins in one hit: if $p=1$, the two states
are completely degenerate, but the random dilution introduces a subextensive
shift in the total energy. In fact, the inversion does not change the
nearest-neighbor energy, but changes the sign of the magnetic part. In the
pure system the magnetic energy of the fully ordered antiferromagnetic state
is zero but, in the presence of random dilution, the number of spins aligned
or misaligned with the field $H$ is a random variable. So, the total magnetic
contribution to energy is of order $\left[p(1-p)N\right]^{1/2}$. It follows
that the $M_\mathrm{s}=-p$ state has an energy shift of order
$2\left[p(1-p)N\right]^{1/2}$ with respect to the $M_\mathrm{s}=p$ one and its
Boltzmann weight is depressed (an analogous effect of degeneration removal is
present in the RFIM).  A ``quasisymmetric'' state may exist if there is a
configuration of spins in which almost all the spins are reversed with
respect to the $M_s=p$ state
such that the sum of the energies of unsatisfied bonds
cancels the magnetic excess. If this is the case, then the two states are
degenerate but the probability distribution of the order parameter results
peaked around asymmetric values.  The magnetic energy excess is a subextensive
effect and is expected to be suppressed as $L$ increases. Nevertheless, the
probability of transitions between states of opposite spontaneous staggered
magnetizations decreases for large system, which is a major problem for
simulations.

\subsection{Observables}\label{OBSERVABLES}

In the following, $\thav{(\cdot \cdot\cdot)}$ denotes thermal averages
(including averages of real replicas) and $\smav{(\cdot\cdot\cdot)}$ indicates
a sample average (average on the disorder). Measures focused on several
observables: the order parameter, i.e., the average value of the
\emph{staggered magnetization}

\be
\label{eg:stag_mag}
\average{M_\mathrm{s}}=\frac{1}{V} \average{\sum_j \epsilon_jS_je^{i\pi(\sum_{\mu=1}^dj_\mu)}}\,,
\ee
($j_\mu$ is the $\mu$-th lattice coordinate of site $j$) whose values are
limited in the interval $-p \leq M_\mathrm{s} \leq p$ (in average for large
lattices); the average energy densities are \bea
\label{eq:energy_tot}
\frac{1}{V}\average{\mathcal{H}} & = & \average{E} =\average{E_\mathrm{K}}+H\average{E_\mathrm{M}}\,, \\
\label{eq:energy_K}
\frac{1}{V}\average{\mathcal{H}_\mathrm{K}} & = & \average{E_\mathrm{K}} = \frac{1}{V} \average{\sum_{<i,j>}\epsilon_iS_i\epsilon_jS_j}\,, \\
\label{eq:energy_M}
\frac{1}{V}\average{\mathcal{H}_\mathrm{M}} & = & \average{E_\mathrm{M}} = -\frac{1}{V}
\average{\sum_i\epsilon_iS_i} \,,
\eea
with $\mathcal{H}_\mathrm{K}$ and $\mathcal{H}_\mathrm{M}$ respectively the kinetic and magnetic
contributions to the Hamiltonian. The definition of $E_\mathrm{M}$ coincides
with the definition of the usual \emph{magnetization density}.

We also computed the average values of the squares and fourth powers of the
above quantities, and some cumulants and susceptibilities: given an observable
$\frac{1}{V}\mathcal{A}=A$ we compute the \emph{Binder cumulant}:
\be
\label{eq:binder}
g_4^A=\frac{1}{2}\left(3-\frac{\average{A^4}}{\smav{{\thav{A^2}}}^2}\right)
\ee
and the connected and disconnected susceptibilities 
\bea
\label{eq:kic}
\chi_\mathrm{c}^A & = & V\smav{\thav{A^2}-{\thav{A}}^2}\,, \\
\label{eq:kidis}
\chi_\mathrm{dis}^A & = & V\smav{{\thav{A}}^2}\,.
\eea
These are the ordinary susceptibilities in case $A=M_\mathrm{s}$, while
$\chi_\mathrm{c}^\mathcal{H}$ is proportional to the specific heat $C_v$.

The lack of $Z_2$ symmetry, explained in Sec.~\ref{MODEL-PHASE-DIAGRAM} makes
mandatory the use of connected correlation functions in finite lattice
sizes, especially in the case of the order parameter $M_\mathrm{s}$.
Yet, the connected \emph{staggered} susceptibility
$\chi_\mathrm{c}^{M_\mathrm{s}} =
V\smav{\thav{{M_\mathrm{s}}^2}-{\thav{{M_\mathrm{s}}}}^2}$ does not show a
peak in the $(T,H)$ ranges we considered, so we also study the behavior of the
connected and disconnected staggered susceptibilities defined with the absolute
value of the staggered magnetization: \bea
\label{eq:kiac}
\chi_\mathrm{c} & = & V\smav{\thav{M_\mathrm{s}^2}-{\thav{\left|M_\mathrm{s}\right|}}^2}\,,\\
\label{eq:kiadis}
\chi_\mathrm{dis} & = &
V\smav{{\thav{\left|M_\mathrm{s}\right|}}^2}\,.  \eea In the
following, when no observable subscript is specified in the
susceptibility symbol, we will be referring to Eqs. (\ref{eq:kiac})
and (\ref{eq:kiadis}).

It will turn out useful to 
 define a correlation length on a finite lattice by the following analogy
with a (lattice) Gaussian model:\cite{cooper} \be
\label{eq:xigen}
\xi^2 = \frac{G(k_1)-G(k_2)}{k_2^2G(k_2)-k_1^2G(k_1)}\,, \ee with
$k^2=4\sum_{\mu=1}^d\sin^2{(k_\mu /2)}$ on a discrete lattice and $G(k)$ the
momentum-dependent propagator \bea
\label{eq:gk}
G(k) & = & V\smav{\thav{F(k)F(-k)}-\thav{F(k)}\thav{F(-k)}}\,, \\
\nonumber & = &\frac{G_0}{\xi^{-2}+k^2}\ \ (k^2 \ll \xi^{-2}) \,,\\
\label{eq:fk}
F(k) & = & \frac{1}{V}\sum_j \epsilon_jS_j
e^{i\sum_{\mu=1}^d(k_\mu + \pi)j_\mu} \,,\\
\nonumber ( G(0) & = & \chi_c^{M_s},\ F(0) =  M_s )\,,
\eea
and $F(k)$ the staggered Fourier transform of the spin field. Also in this
case we use the connected part for $G(k)$. 
Choosing $k_1=(0,0,0)$ and $k_2=(2\pi/L)\hat{k}_\mu$ as one of the minimum
wave vectors ($\hat{k}_\mu,\ \mu=1,\dots,d$ are the $d$ versors in the reciprocal space), we have
\be
\label{eq:xi}
\xi^2 = \frac{1}{4\sin^2(\pi/L)}\left(\frac{\chi_c^{M_s}}{\left[\sum_{\mu=1}^d
  G\left({(2\pi/L)\hat{k}_\mu}\right)\right]/d}\right)\ \ .
\ee
Equation (\ref{eq:xi}) is a good estimate of the correlation length only in the
disordered phase but is useful to identify the critical region where $\xi/L$
should be a nontrivial universal value.

Finally, with mass storage not being a problem on modern equipment, it is easy to
compute derivatives with respect to inverse temperature $\beta=1/T$ and applied
field $H$, through connected correlations.  In particular, the specific heat
is
\be
\label{eq:cv}
C_v=\frac{1}{V}\smav{\frac{d\thav{\mathcal{H}}}{dT}}\ \ \ .
\ee

\subsection{Finite-size scaling in second-order phase transitions}\label{FSS-2-SUBSECT}

We made use of finite size scaling,~\cite{barber} both studying the
behavior of peaks of susceptibilities and applying the quotient method
(QM)~\cite{quotient} to extract values for critical exponents.  Let us briefly
recall both.

Consider an observable $A$, that in the infinite volume limit behaves
as $(T-T_c)^{-a}=t^{-a}$ near the critical region ($t$ is the reduced
temperature). Then, disregarding correction-to-scaling terms, we
expect the following temperature dependency on a finite lattice of
size $L$ \be
\label{eq:scal}
A(L,t)=L^{a/\nu}f_A(tL^{1/\nu})\,,
\ee 
with $\nu$ the correlation length exponent, $\xi\propto t^{-\nu}$,
and $f_A(s)$ a smooth universal scaling function showing a peak at some value $s_m=t_m(L)L^{1/\nu}$. It follows
that $T_m(L)-T_c^{\infty} \propto L^{-1/\nu}$. In addition, the scaling of the
peak-height gives the value of the critical exponent $a$.

The QM is based on the same scaling ansatz:
\be
\label{eq:scal_q}
A(L,t)=L^{a/\nu}g_A(\xi^{-1}(L,t)L) \ee We compare data in two lattices $L_1$
and $L_2$ at the (unique) reduced temperature $t^*$ where the
correlation-length in units of the lattice size coincides,
$\xi(L_1,t)/L_1=\xi(L_2,t)/L_2$. At this temperature we have, apart from
corrections to scaling: \be
\label{eq:quotient}
\frac{A(L_1,t^*)}{A(L_2,t^*)}=\left(\frac{L_1}{L_2}\right)^{a/\nu} \ee 
Note
that the {\em crossing temperature} $T^*(L_1;L_2)$ approaches the critical
temperature for large $L$ much faster than the peak of any susceptibility:
$t^*=T^*(L_1;L_2)-T_c^{\infty} \propto L^{-\omega-1/\nu}$ ($\omega$ is the
leading correction-to-scaling exponent).

From the definition [Eq. (\ref{eq:xi})] of the correlation length, one
sees that, respectively, $\xi/L\sim O(L^{cd})$ in the ``ordered''
(low $T$, low $H$) phase and $\xi/L\sim O(1/L)$ in the ``disordered''
phase.  The constant $c$ is $1/2$ in the case when the ordered phase
has a $Z_2$ global symmetry, for in finite lattices the disconnected
susceptibility would vanish.  Near a second-order transition, $\xi/L$
does not depend on $L$, so there is a region in which $\xi/L$ for
different lattice sizes cross each other.  The method then consist in
finding the value $T^*(L_1;L_2)$ at which this crossing happens and
extracting the exponent $a/\nu$ by means of Eq. (\ref{eq:quotient}).

We apply the methods to several observables to extract exponents, in particular
\bea
\label{eq:scal_chic}
\chi_\mathrm{c}^{M_\mathrm{s}},\chi_\mathrm{c} & \longrightarrow & a=\gamma=\nu(2-\eta)\,,\\
\label{eq:scal_chidis}
\chi_\mathrm{dis}^{M_\mathrm{s}}, \chi_\mathrm{dis} & \longrightarrow & a=\overline{\gamma}=\nu(2-\overline{\eta})\,,\\
\label{eq:scal_cv}
C_v & \longrightarrow & a=\alpha\,,\\
\label{eq:scal_ms}
\left|M_\mathrm{s}\right| & \longrightarrow & a=-\beta\,,\\
\label{eq:scal_dbxi}
\partial_\beta\xi & \longrightarrow & a=1+\nu \,.
\eea
Notice that we follow Ogielski and Huse~\cite{ogielski} in defining $\overline\eta$.

\subsection{Finite-size scaling for first-order phase transitions}\label{FSS-1-SUBSECT}
Finite-size effects in first-order phase transitions on {\em pure
systems} are qualitatively similar to their second-order counterpart,
provided that one considers effective critical
exponents.~\cite{binder_rev,binlau_I,binlau_II} With the assumption
that the lattice size is much larger than the correlation length,
simple scaling relations hold for the size of the broadened
transition region, the height of the peak of the specific heat and the
extremal point of the binder cumulant for the energy density: denoting
with subscripts $+$ and $-$ values of observables of the two competing
phases at a first-order transition in the infinite volume limit (one
of the phases can be degenerate) and being $Q=E_{+}-E_{-}$ the
\emph{latent heat}, one has the following in a finite system:~\cite{binlau_II} \bea
\label{eq:scal_I_T}
T^*(L)-T_c & = & a(Q)L^{-d}\,,\\
\label{eq:scal_I_cv}
C_v(T^*) & = & c_1(C_{v+},C_{v-})+c_2(Q)L^{d}\,,\\
\label{eq:scal_I_ge}
1-g_4^E(T^*) & = & g_1(E_{+},E_{-}) \\ \nonumber & + & g_2(E_{+},E_{-},C_{v+},C_{v-})L^{-d}\,,
\eea
where, in particular, $a(Q)$, $c_2(Q)$ and $g_1(E_{+},E_{-})$ vanish if the latent heat
is zero, i.e., $E{+}=E_{-}$. $C_{v\pm}$ are the specific heats of the
$\pm$ phases. Finally, the susceptibility also diverges with the
volume of the system.

However, in the presence of disorder the scaling
law of $T^*(L)-T_c$ should be modified (see the appendix)
\begin{equation}
\label{eq:scal_mod_T}
T^*(L)-T_c  =  b(Q)L^{-d/2}\,.
\end{equation}
This follows, for instance, from a simple mean-field
argument,~\cite{Chatelain01} that yields a linear relation between
the critical temperature and the number of spins in the
samples. Since the average spin density fluctuates as $L^{-d/2}$, we
expect this to be the width of the critical region on finite lattices.
Furthermore, the specific heat and the connected susceptibility may
diverge only as fast as $L^{d/2}$. See the appendix for a detailed
discussion of these bounds.

Hence,  assuming that the observables diverge as much as
possible, we can write the following ``apparent'' critical exponents
for a disordered first-order transition:
\be
\frac{1}{\nu}=\frac{d}{2}\ ,
\ee
\be
\frac{\alpha}{\nu}=\frac{d}{2}\ ,
\ee
\be
\frac{\gamma}{\nu}=\frac{d}{2}\ .
\ee
From the last equation and using $\eta=2-\gamma/\nu=2-d/2$, we find  in
$d=3$ that $\eta=0.5$ and $\nu=2/3$.

If we assume that the averaged probability distribution of the energy
$\overline{\langle P(E)\rangle}$ is composed (in the thermodynamic limit and
at the critical point) bof the sum of Dirac deltas, we should obtain a divergence
$L^d$ for the normalized variance of this averaged probability, obtaining
(e.g., for the energy) \be L^d \left(\overline{\langle E^2 \rangle}-
  \overline{\langle E \rangle}^2\right)= Q^2 L^d \,.  \ee In particular, we
should recover Eq. (\ref{eq:scal_I_ge}) for $g_4^E$ which is computed with
$\overline{\langle P(E)\rangle}$ [see Eq. (\ref{eq:binder})].  Please note that
the width of $\overline{\langle P(E)\rangle}$ is {\em not} related to the
specific heat, which is rather related to $L^d \left(\overline{\langle E^2
    \rangle}- \overline{\langle E \rangle^2}\right)$.

\begin{figure}[t]
\includegraphics[width=\columnwidth]{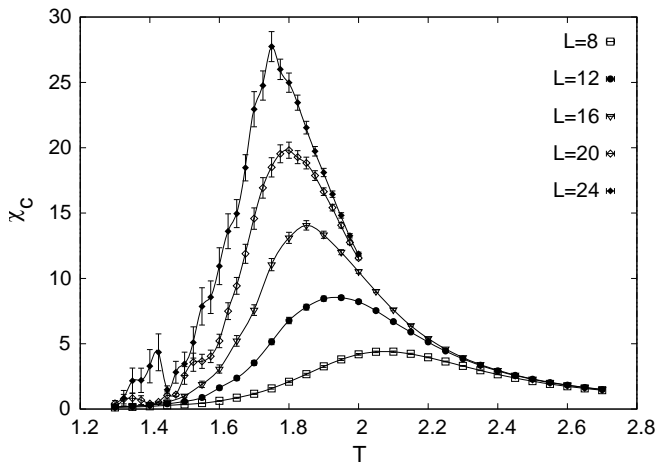}
\caption{The connected susceptibility computed with $\left|M_s\right|$ as
  function of $T$ for lattice sizes $L=8,12,16,20$ and $24$. Lines are interpolating
  splines as a guide to the eye.}
\label{fig:chi}
\end{figure}
\begin{figure}[h]
\includegraphics[width=\columnwidth]{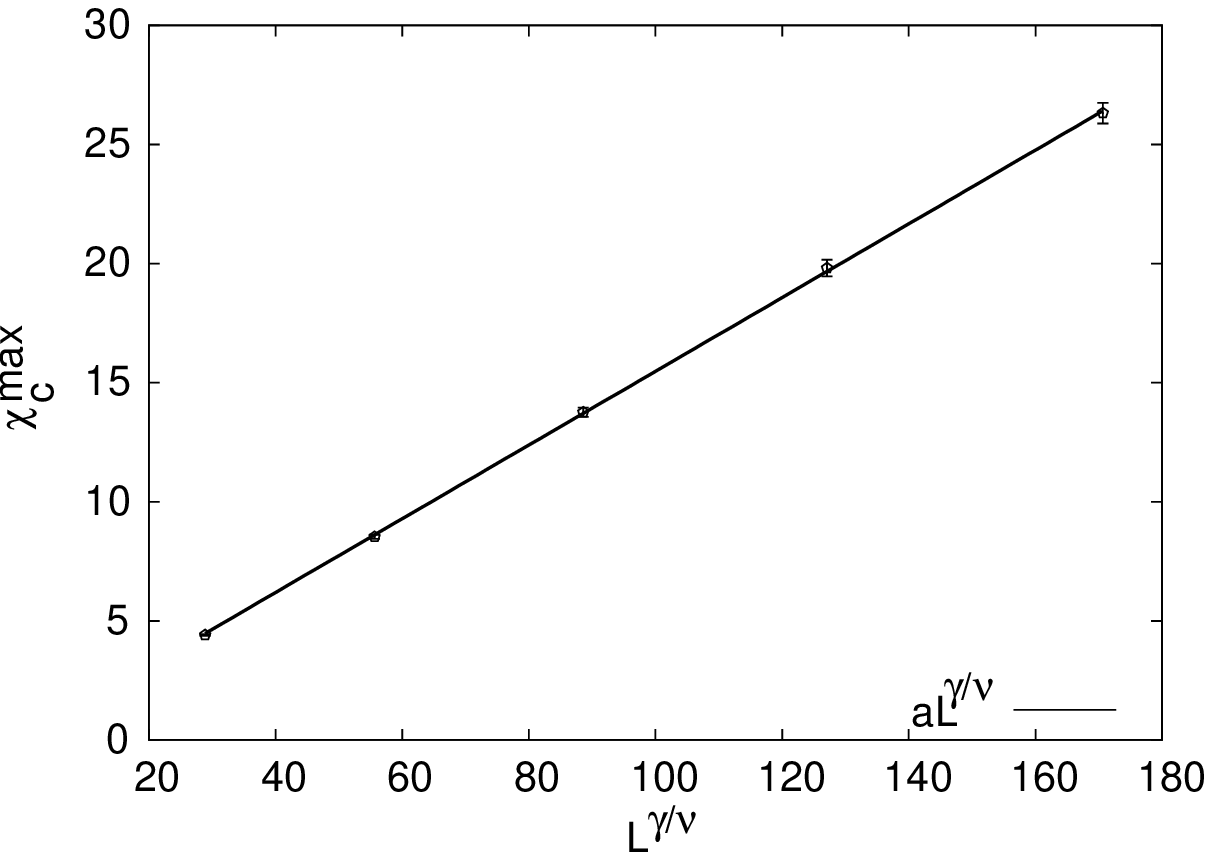}
\includegraphics[width=\columnwidth]{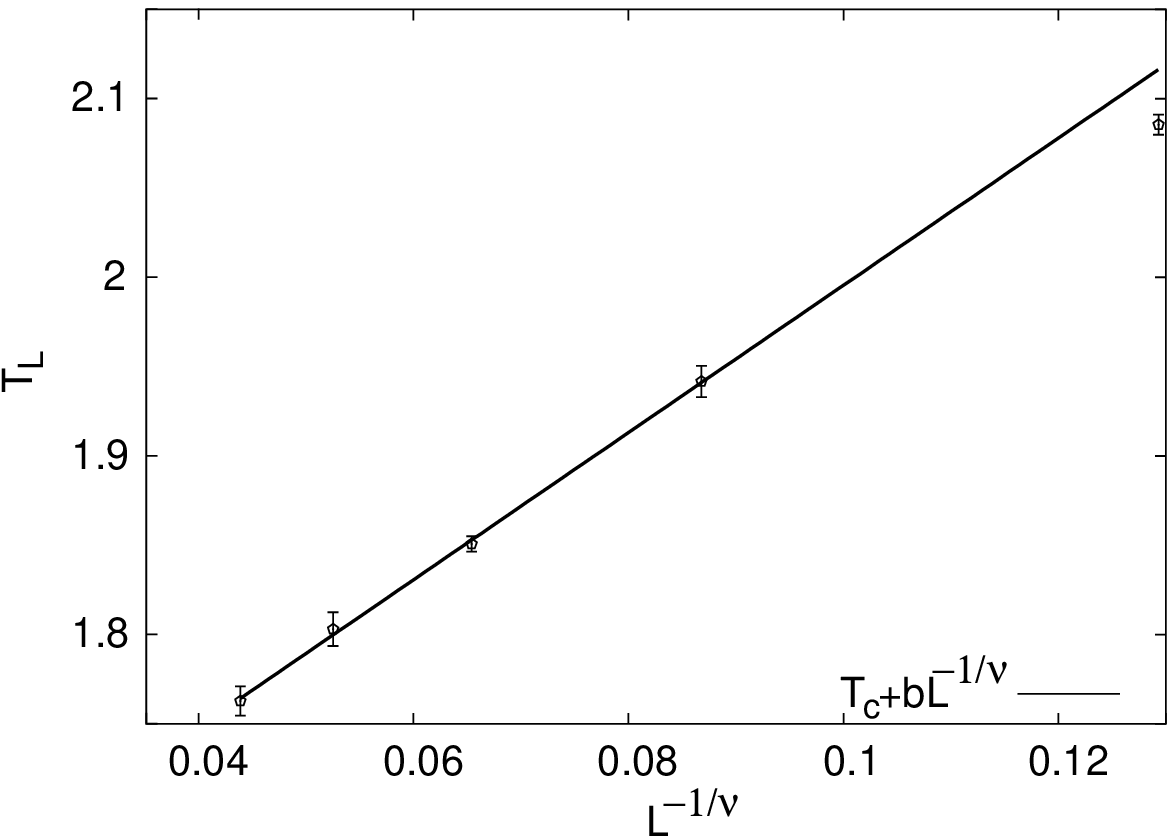}
\caption{Top: the susceptibility peak height as function of
  $L^{\gamma/\nu}$. Bottom: Peak position as function of $L^{-1/\nu}$. Solid
  lines are the fitting power-law functions. See section~\ref{NUMERICAL-1} for more details.}
\label{fig:chi_pk}
\end{figure}

\section{Simulation details}\label{SIMULATION}

We simulate the model using the usual Metropolis algorithm with
sequential spin flip schedule and the parallel tempering
technique.~\cite{pt} We restricted our simulation to the \be
H=1.5T\label{diagonal} \ee diagonal in the $(T,H)$ plane in order to
keep away from the crossover to the zero field case.  This should also
avoid problems with an oblique crossing of the transition line and
will help in the comparison with previous numerical
simulations.~\cite{ogielski}

The critical temperature on this diagonal stays around $T_c=1.5$, and
we simulated $N_T=29$ temperatures for every lattice size at equally
spaced $(T,H)$ values along this diagonal. For smaller lattices
($L=8,\ 12,$ and $16$) the $T$ values were in the range $[1.3,2.7]$, while
for the larger sizes ($L=20$ and $24$) the temperature range was
$[1.3,2.0]$. For every lattice size, $1280$ samples (different
realizations of the disorder) were simulated. Statistics is also
doubled as our program processes two real replicas per sample, with
the same disorder, at each $(T,H)$ value.

\begin{table}[t]
\begin{tabular}{||l|c|c|c|c|c||}
\hline
$L$ & $N_\mathrm{samples}$ & $N_T$ & $t_\mathrm{sim}$ & $T_\mathrm{max}$ & $T_\mathrm{min}$ \\
\hline
\hline
$8$ & $1280$ & $29$ & $2\times 10^6$ & $2.7$ & $1.3$ \\   
\hline
$12$ & $1280$ & $29$ & $2\times 10^6$ & $2.7$ & $1.3$ \\   
\hline
$16$ & $1280$ & $29$ & $8\times 10^6$ & $2.7$ & $1.3$ \\   
\hline
$20$ & $1280$ & $29$ & $1.6\times 10^7$ & $2.0$ & $1.3$ \\   
\hline
$24$ & $1280$ & $29$ & $2.4\times 10^7$ & $2.0$ & $1.3$ \\   
\hline
\end{tabular}
\caption{Parameters characterizing the simulation. See text for details.}
\label{tab:sim}
\end{table}

The use of optimized asynchronous multispin coded update routines in
our programs allowed us to thermalize systems on lattices with size up
to $L=24$.  The program can perform Metropolis update at a
$1.3~\mbox{ns}/\mbox{spin}$ rate on a conventional $64$ bits Intel CPU
at 3.4 GHz. Of course, the use of parallel tempering (PT) slows down
the performance of the multispin code simulation, 
but we can limit the loss of performance if we let
the program perform PT swaps every many Metropolis lattice sweeps. We
verified that a PT swap trial every 20 Metropolis lattice sweeps also
allows for hotter replicas to decorrelate before the exchange with
colder ones, at the cost of a factor of $1.5$ in performances. Cluster
update algorithms did not prove convenient due to a dramatic increase in
total computational load. In what follows, we consider simulation time
units such that $20$ Monte Carlo (MC) steps are $20$ Metropolis full
lattice updates plus $1$ PT step.

Simulating $1280$ samples for the largest lattice ($L=24$ and
$24\times 10^6$ MC steps) took about four weeks and 20 computation
nodes on the Linux cluster at BIFI.  By monitoring nonlocal
observables (like the susceptibilities), we have checked that the runs are
thermalized: we have reached a plateau in all the nonlocal
observables we are measuring. In particular, for each lattice size, let
$t_\mathrm{sim}$ be the total time in MC steps devoted to simulate a
sample, the time needed to achieve equilibrium always resulted shorter
than $t_\mathrm{sim}/2$. Indeed, we discarded measures at all times
$t<t_\mathrm{sim}/2$.  Simulation parameters are summarized in Table
\ref{tab:sim}.

Two further thermalization test were provided by the parallel
tempering statistics: (1) we have checked that the temperature samples
perform all the road from higher to lower temperatures and come back;
(2) the temperature samples have stayed essentially the same Monte
Carlo time in all the temperatures simulated.

\section{NUMERICAL RESULTS}\label{NUMERICAL}

\subsection{Second-order phase transition scenario}\label{NUMERICAL-1}

We will use first the old fashioned peak method and turn later to the
quotient method. In this way we will obtain complementary information.

In Fig. \ref{fig:chi} we show the staggered magnetization connected
susceptibility $\chi_\mathrm{c}$ data. Clear peaks are present from which
it is
possible to extract information on exponents $\gamma$, $\nu$ and $\eta$. There
is a lot of noise in the signal for $\chi_\mathrm{c}$ at low temperatures for
large lattice sizes ($L=20$ and $L=24$). This is almost exclusively due to the
disconnected part of the connected susceptibility, which is difficult to
obtain because of metastability (see Sec.~\ref{NUMERICAL-3}).

The exact peak position and height are located
by means of cubic polynomial interpolation, and by using the standard
second-order phase transition equations for the peak and the
position of the maximum of the susceptibility ($\chi_{\mathrm{max}}
\propto L^{\gamma/\nu}$ and $T_c-T_\mathrm{max} \propto L^{-1/\nu}$),  we obtain
\be
\label{eq:ki_c_eta}
\frac{\gamma}{\nu}=(2-\eta)=1.6(1) \rightarrow \eta=0.4(1)\, ,
\ee 
\be
\label{eq:ki_c_nu}
\nu=1.0(3)\, ,
\ee 
\be
\label{eq:ki_c_Tc}
T_c=1.58(8)\, , \ee (data for $L=8$ has been excluded in determining $T_c$ and
$\nu$).  These results are fully consistent with $\beta=0$, even if we are
using a second-order ansatz in the analysis.
 
Figure \ref{fig:chi_pk} shows the dependence of peak heights and
positions on the size $L$.  These estimates are compatible with
previous ones by Ogielski and Huse:~\cite{ogielski} $T_c=1.50(15)$,
$\nu=1.3(3)$, and $\eta=0.5(1)$.~\cite{fn2}
Ground states calculations by Hartmann and Nowak~\cite{hartmann1} for
the DAFF (but at dilution $p=0.55$) gave $\nu=1.14(10)$.

\begin{figure}[h]
\includegraphics[width=\columnwidth,trim=12 15 18 0]{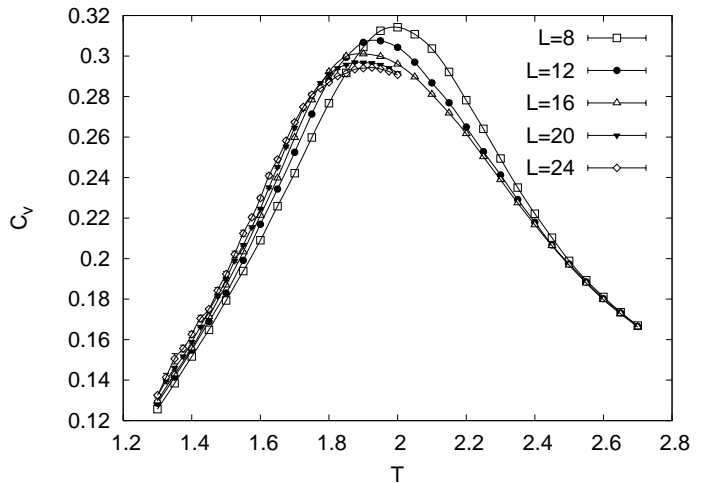}
\caption{The specific heat as function of $T$ for various lattice sizes. The
  main feature is in the decrease and flattening of the peaks as $L$ increases.}
\label{fig:cv}
\end{figure}

The specific heat shows no tendency to diverge at all near the
transition region. On the contrary, the peak of $C_v$ tends to
slightly decrease and broaden as system size increases (see
Fig. \ref{fig:cv}).  This is probably an artifact due to the large
slope of the path in the $(T,H)$ plane that we simulated [see
Eq.(\ref{diagonal})]. In fact, the larger the system size, the lower
the peak height. However, note that the specific heat has a contribution
(from the magnetic energy) with an explicit linear dependence on the
field strength (also the magnetization depends strongly on
it). Anyhow, this supports a scenario of negative (maybe vanishing)
$\alpha$, as reported, for example, by Rieger and Young \cite{rieger_II},
Rieger\cite{rieger_I}, Middleton and Fisher\cite{middleton} in their simulation of the random field
Ising model and in experiments.~\cite{belanger}  We shall discuss
further the specific heat in the following.  At the time being, note
that the peak
position of $C_v$ may be fitted to the usual power law and we find \be
\label{eq:cv_Tc}
T_c=1.68(4)\,, \ee which is compatible with the estimate given in
Eq. (\ref{eq:ki_c_Tc}).  This fit provides no information on the $\nu$
exponent ($\nu=2(2)$).

\begin{figure}[b]
\includegraphics[width=\columnwidth,trim=12 15 18 0]{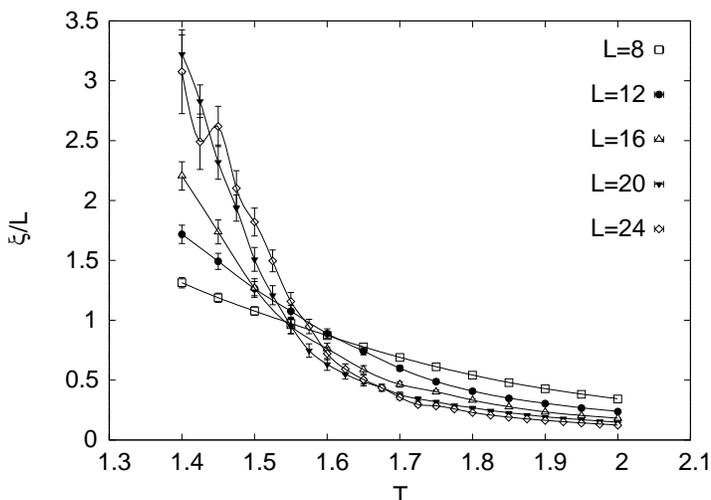}
\caption{The cumulant $\xi/L$ as function of the temperature for various
  system sizes. Lines are interpolating splines and serve only as a guide to
  the eye. Note the noise in the curve for $L=24$ in all the crossing region.}
\label{fig:xic}
\end{figure}

We can extract further information on several exponents by means of
the QM. Figure  \ref{fig:xic} shows a clear crossing of the ratio
$\xi/L$ as function of $T$ for different values of $L$.  Data for
$L=20$ and $L=24$ are quite noisy, again due to difficulties in
measuring the disconnected part of $G(k)$ [Eq.  (\ref{eq:gk})] but
still allow for locating a crossing temperature with other curves. As
expected on general grounds, the crossing temperatures stay well away
from the positions of the peak of $C_v$ and $\chi_\mathrm{c}$ but lie
fairly close to the $T_\mathrm{c}^\infty$ value that has been extrapolated
from the peaks position. Indeed, in the absence of scaling
corrections, there should be no system size dependency of the crossing
temperature. Such dependency, if any, carries important information on
scaling corrections.~\cite{quotient} Unfortunately, in our case, there
is no clear systematic dependence of $T_\mathrm{cross}$ on the
lattice size as, for example, for small systems $T_\mathrm{cross}$
tends to shift to lower values as $L_1$ and $L_2$ increase, while it is
sensibly shifted toward higher values when lattice size $L=20$ or
$L=24$ is considered. This probably indicates that a crossover to
first-order behavior is showing up.

We show in Table \ref{tab:exps} our results for values of exponents $\alpha$,
$\beta$, $\eta$ and $\overline{\eta}$, obtained from the QM. Unfortunately, we
have not been able to measure $\partial_\beta \xi$ with enough precision to
give a direct estimate for the thermal exponent $\nu$.
\begin{table}[t]
\begin{tabular}{|c|c|c|c|c|c|c|}
\hline \hline
$L_1,L_2$ &  $T_\mathrm{cross}$  &     $\eta$    & $\overline{\eta}$ &   $\alpha/\nu$   &       $\beta/\nu$\\
\hline \hline
8,12    & 1.6(2)  &   0.5(1)    &   -1.0(1)  &   0.091(6)   &   0.07(6)  \\
8,16    & 1.54(6) &   0.8(2)    &   -0.99(3) &   0.07(2)   &   0.05(2)  \\
8,20    & 1.55(2) &   0.4(2)    &   -0.97(1) &   0.070(1)   &   0.049(5) \\
8,24    & 1.58(2) &   0.2(2)    &   -0.94(1) &   0.083(1)   &   0.06(1)  \\
12,16   & 1.5(1)  &   1.1(6)    &   -1.00(3) &   0.07(2)   &   0.04(3)  \\
12,20   & 1.53(4) &   0.1(4)    &   -0.95(3) &   0.08(1)   &   0.05(1)  \\
12,24   & 1.57(4) &   0.0(3)    &   -0.92(2) &   0.087(7)   &   0.06(1)  \\
16,20   & 1.55(5) &   -0.5(8)   &   -0.90(4) &   0.09(2)    &   0.056(4) \\
16,24   & 1.59(4) &   -0.4(5)   &   -0.85(2) &   0.09(1)   &   0.08(2)  \\
\hline \hline
\end{tabular}
\caption{Exponents and crossing temperatures extracted with the QM applied to the intersection of the
  cumulant $\xi/L$.}
\label{tab:exps}
\end{table}

Maybe the most striking result in Table \ref{tab:exps} is the
smallness of the $\beta$ exponent, indicating that the order parameter
could be discontinuous at the transition. A similar behavior has been
found for the RFIM \cite{rieger_I,hartmann1} and in experiments.~\cite{belanger}
A very small (but definitely positive) value of $\beta$ has been found
also by Falicov \emph{et al.}~\cite{falicov} by means of
renormalization-group calculations for the binary RFIM. They also
calculated magnetization curves as functions of the temperature and
field strength, showing abrupt jumps at the transition point.

The value of $\overline{\eta}$ agrees with the one found in Ref.
[\onlinecite{ogielski}],$\overline{\eta}=-1.0(3)$ and agrees with the
smallness of the order parameter exponent and the estimate [Eq.
(\ref{eq:ki_c_nu})] of $\nu$ as the relation
$\beta=(1+\overline{\eta})\nu/2$ holds.  Note also that, given the
value  of $\eta$ from Eq. (\ref{eq:ki_c_eta}), the
Schwartz-Soffer\cite{schwartz} relations $2(\eta-1)\geq \overline{\eta}
\geq -1$ are satisfied as equalities within errors.  We see that, at
larger sizes, the value of $\eta$ decreases down even to negative
values (showing a large error) but is always compatible with our previous
estimate (at least at two standard deviations): $\eta= 0.4(1)$. Also
the disconnected susceptibility diverges as $L^3$ (since
$\overline{\eta}$ is very close to $-1$, because $\chi_\mathrm{dis}
\propto L^{2-\overline{\eta}}$).  As for the specific heat exponent,
we know from the Harris criterion~\cite{Harris} that $\alpha$ should
be negative or zero in a disordered second-order transition framework.
Here we report values which are small but definitely positive. It is
also true that if the specific heat had a cusplike singularity, our
data would not allow to evaluate the asymptotic value for $C_v$, and
the estimates for $\alpha$ would be meaningless.

\begin{figure}[h]
\includegraphics[width=\columnwidth,trim=12 15 18 0]{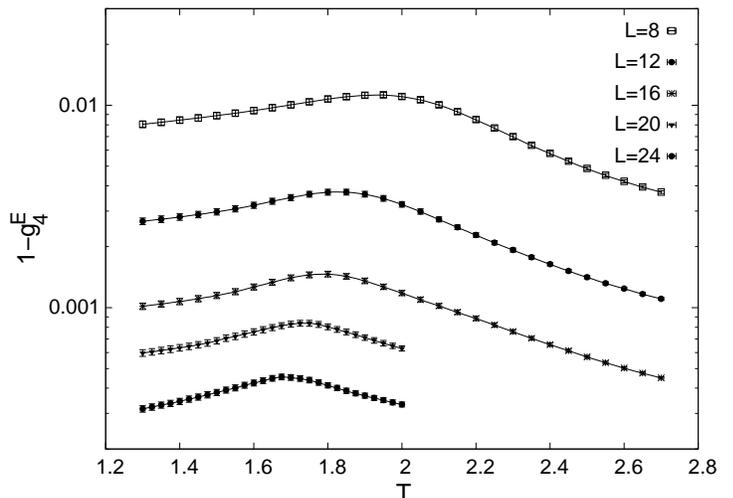}
\caption{The energy Binder cumulant as function of $T$ for all lattice sizes.}
\label{fig:g4e}
\end{figure}
\begin{figure}[t]
\includegraphics[width=\columnwidth]{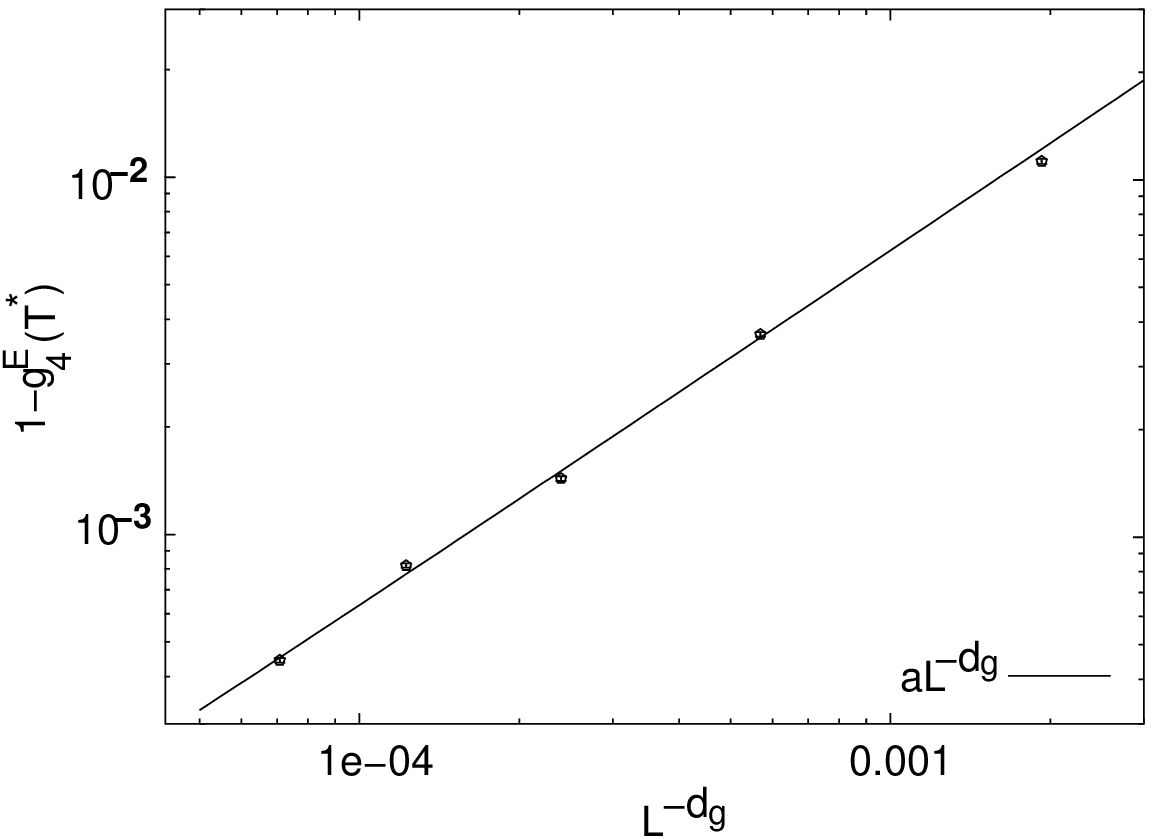}
\includegraphics[width=\columnwidth]{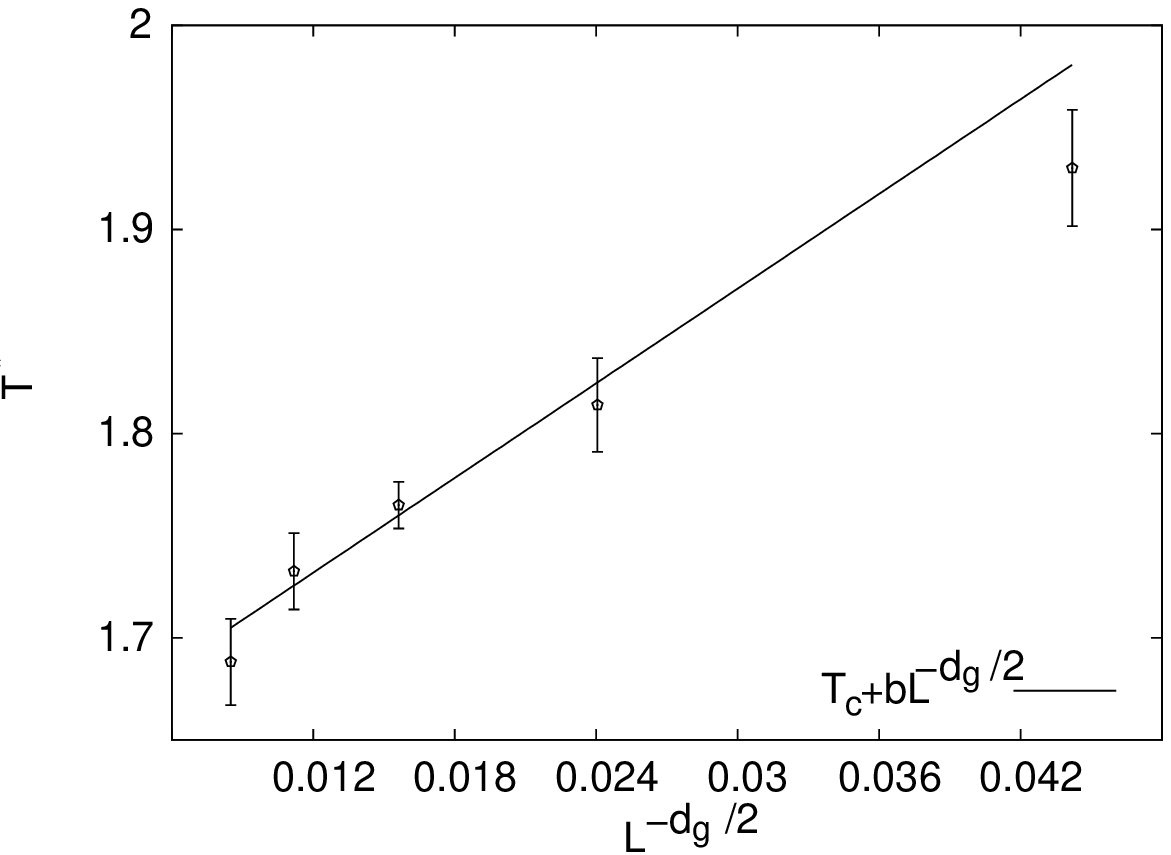}
\caption{Scaling of the minima of Binder cumulant Eq. (\ref{eq:binderE}) (top)
  and of the minima positions (bottom). Fitting functions are also
  showed. In the bottom we have fixed $d_g/2$ to 3/2 in the fit. See
  section~\ref{NUMERICAL-2} for more details.}
\label{fig:g4scal}
\end{figure}

\subsection{First-order phase transition scenario}\label{NUMERICAL-2}

The analysis presented above, based on the hypothesis that a second-order
transition is taking place, looks inconclusive. This is especially clear from
exponent $\eta$, which lies so near to our prediction for a first-order phase
transition: $\eta=0.5$. In addition this exponent in the Schwartz-Soffer
inequality fixes the value of $\overline{\eta}$ to -1, and all our estimates of
the $\overline{\eta}$ exponent are compatible with this value. Of course,
this could be due to finite $L$ corrections to scaling, but we think that the
phase transition is truly first-order.

We now proceed to show that our data are compatible with a \emph{weak
first-order transition} with a very large, but not diverging,
correlation length at the transition point.  A good observable to test
is the Binder cumulant for the total energy density:\cite{binlau_II}
\be
\label{eq:binderE}
g_4^E=\frac{1}{2}\left(3-\frac{\average{E^4}}{\smav{{\thav{E^2}}}^2}\right)\,,
\ee
which is usually easy to measure in simulations because of the good
noise-to-signal ratio  of the energy density. Notice that this Binder
cumulant works directly on the averaged probability distribution of
the energy.
In both the disordered and ordered phases, well away from the transition
temperature, the probability distribution of the energy
$\overline{\langle P(E) \rangle}$ tends to a single
delta function in the thermodynamic limit, so that $g_4^E\rightarrow 1$. In
the case of a second-order transition this is also true at $T_c$, while in
the presence of a first-order transition, we have an energy distribution with more
than one sharp peak, so the infinite volume limit of $g_4^E$ is nontrivial.
Challa \emph{et al.}\cite{binlau_II} obtained the expression for the
nontrivial limit and finite size correction to leading order in the framework of
a double Gaussian approximation for the multipeaked $\overline{\langle  P(E) \rangle}$:
\bea
\label{eq:g4_A}
1-g_4^E(T^*) & = & g_1(E_{+},E_{-}) \\ \nonumber & + & g_2(E_{+},E_{-},C_{v+},C_{v-})L^{-d}\,,\\
\label{eq:g4_res}
g_1(E_{+},E_{-}) & = & \frac{E_+^4+E_-^4}{\left(E_+^2+E_-^2\right)^2}-\frac{1}{2}\,,
\eea
where $T^*$ is the temperature at which the minimum (maximum of $1-g_4^E$)
appears and $g_2$ is a complicated combination of the specific heats and energies of
the infinite volume coexistent states ($+,-$). The term $g_1$ is vanishing if
the latent heat $Q=E_+-E_-$ is zero.

\begin{figure}[b]
\includegraphics[width=\columnwidth]{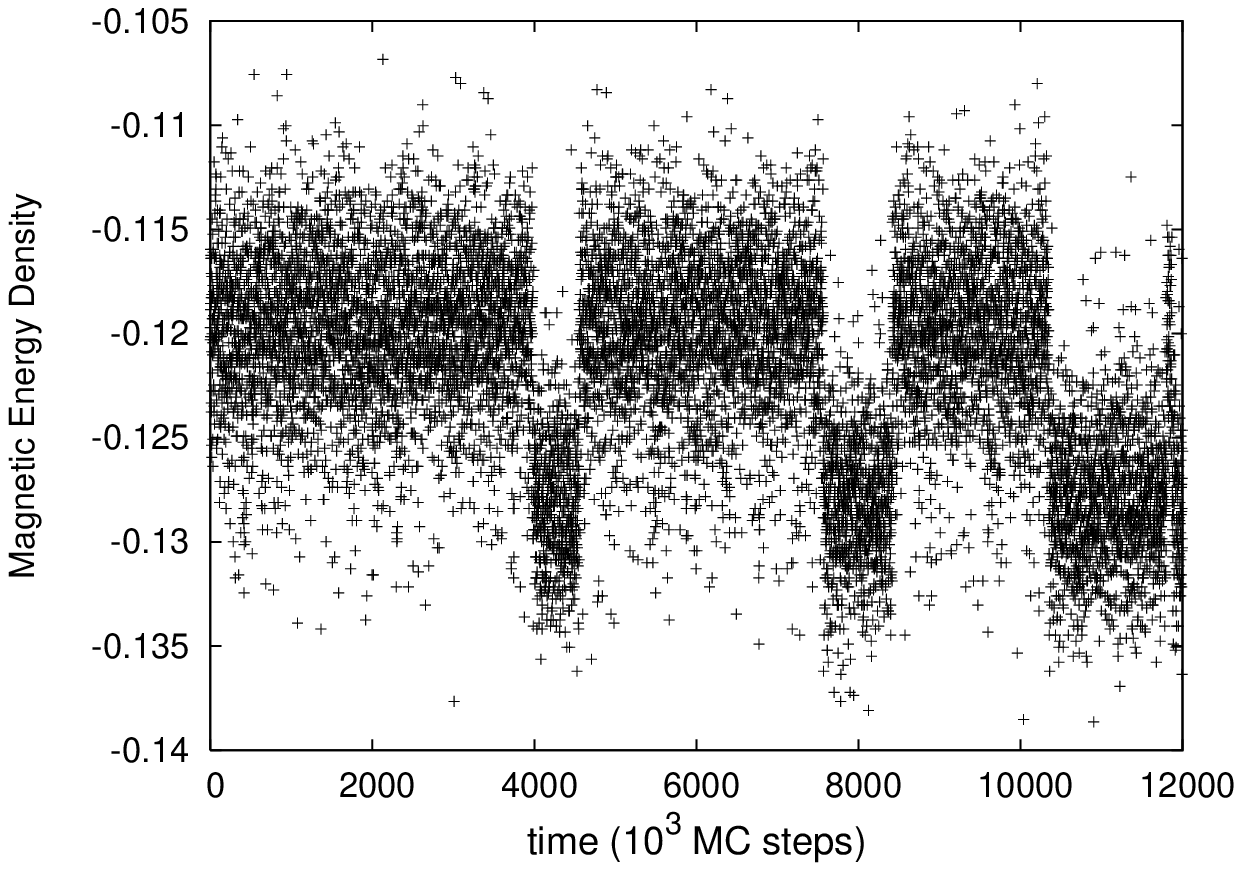}\\
\includegraphics[width=\columnwidth]{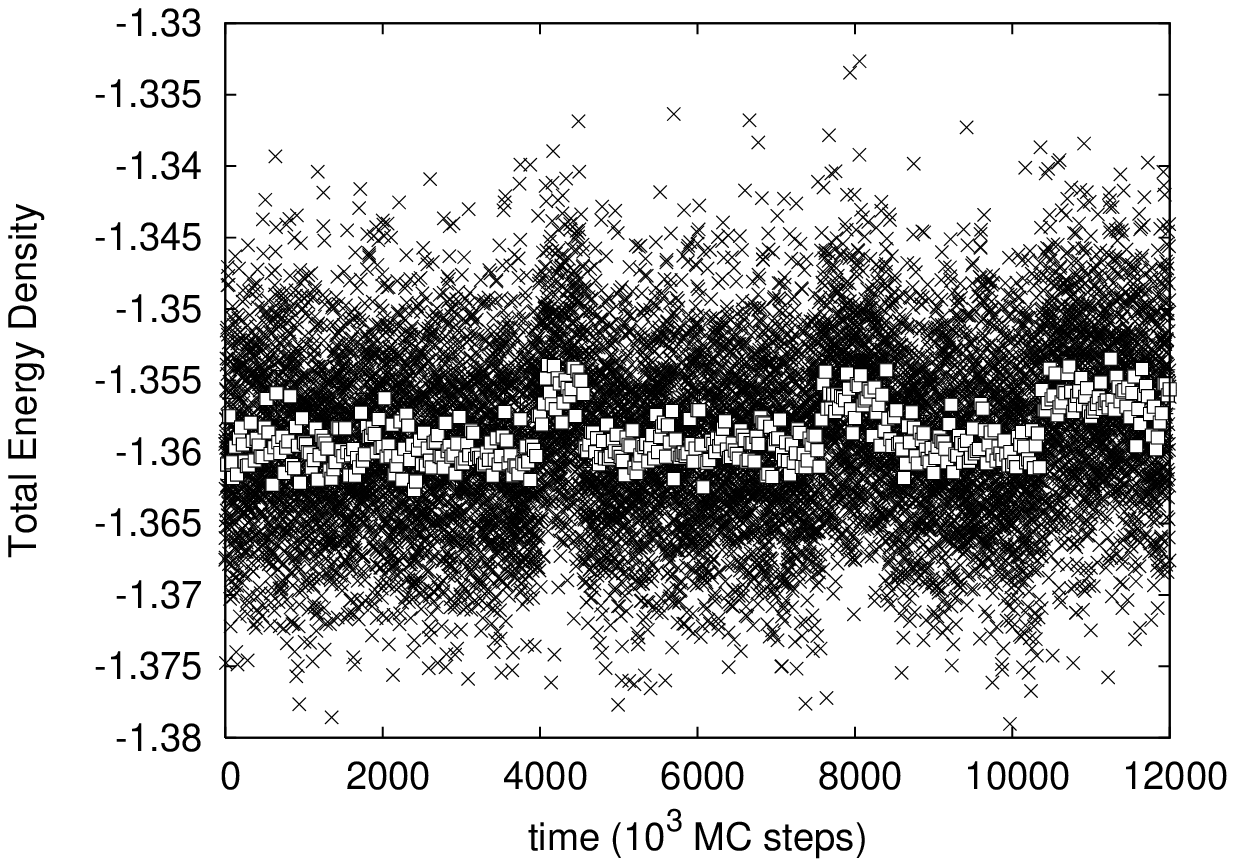}\\
\includegraphics[width=\columnwidth]{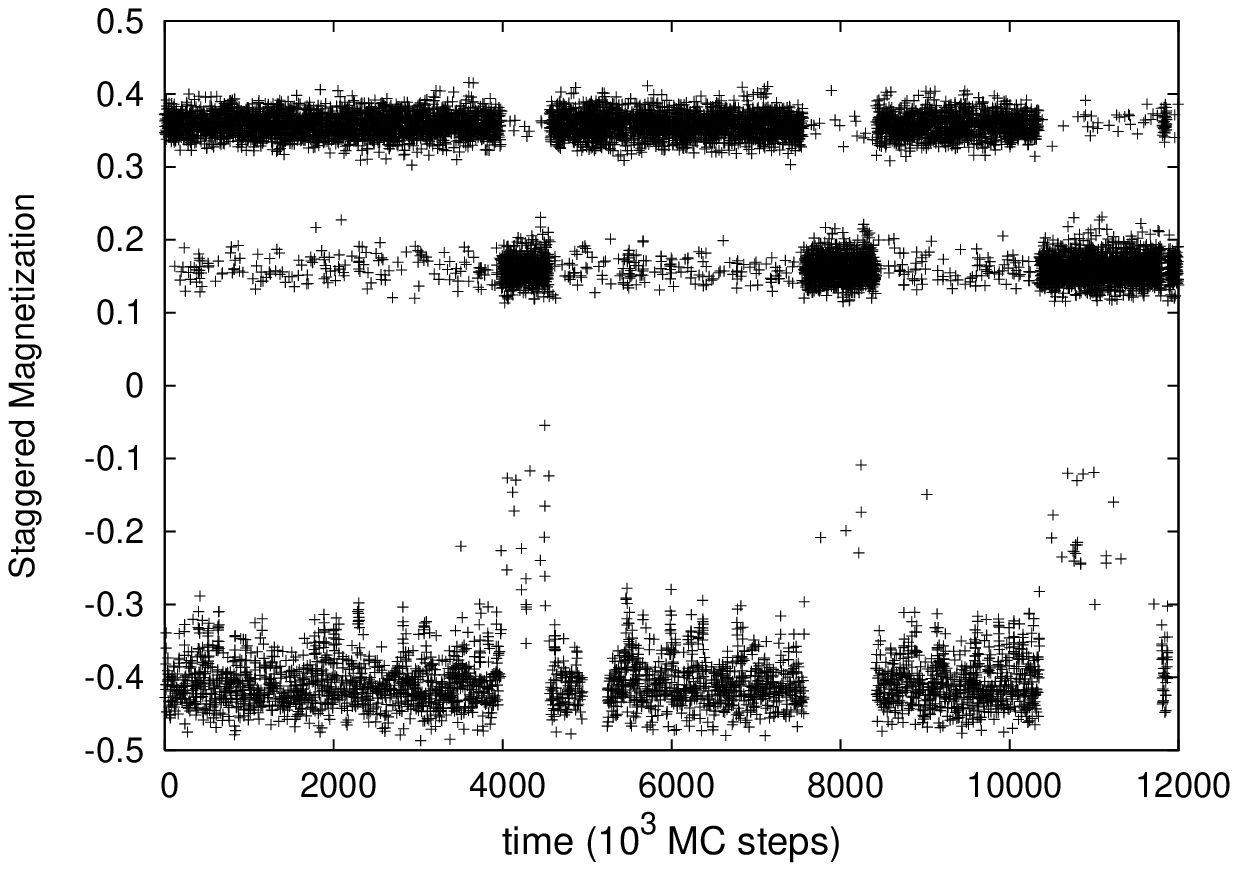}\\
\caption{Monte Carlo histories of magnetic energy density (top), total
  energy density (middle) and staggered magnetization (bottom) for a
  typical sample of lattice size $L=24$ showing jumps between states
  at the transition temperature. The latent heat is clearer when
  binning 25 consecutive Monte Carlo measurements for the total energy
  (open symbols).}
\label{fig:sam163}
\end{figure}

Our data for $1-g_4^E$ effectively show broad peaks at temperatures
$T^*(L)$ shifting toward lower temperature values as $L$ increases
(Fig. \ref{fig:g4e}). We also expect from Eq.~(\ref{eq:scal_mod_T})
that the critical temperature shift should scale as $T^*(L)-T_c\propto
L^{-d/2}$.  

The matching of the data with this model is impressive, (see  Fig. \ref{fig:g4scal}). A
power-law fit against $1-g_4^E(T^*(L))=g_1+g_2L^{-d_g}$ gives \bea
\label{eq:fit_g1}
g_1 & = & 0(2)\times 10^{-4} \,,\\
\label{eq:fit_dg}
d_g & = & 3.0(1)\,, \eea 
where 
 we excluded in the fit the  $L=8$  data [fitting
also $L=8$ data brings $d_g=2.8(1)$ but $g_1$, even if compatible with
zero, has an unphysical negative value]. The extrapolated
transition temperature is (assuming a power $d_g/2=1.5$ and $L>8$) \be
\label{eq:Tc_g4}
T_c=1.64(2) \, .  \ee with a reasonable $\chi^2/\mathrm{DOF}=0.6$.
We also  recall that susceptibility data gave $1/\nu=1.0(3)$ which is 
acceptable: an exponent $3/2$ is within two
standard deviations.

However, the infinite volume limit of $1-g_4^E \left[T^*(L=\infty)\right]$ is
very small, suggesting a zero latent heat for the
transition. Actually, we will show below that one can estimate from
$g_1$ the order of magnitude of the latent heat, which will turn out to
be in agreement with metastability estimates.

\subsection{Metastability}\label{NUMERICAL-3}

A very small latent heat may be very hard to detect due to large thermal and
sample-to-sample (disorder) fluctuations. If this is the case, it should be
possible to detect the latent heat by exploring the behavior of single samples.

Indeed, for our largest systems ($L=24$), around 20\% of the samples
started to display metastability between a disordered, small
$M_\mathrm{s}$ state and a large $M_\mathrm{s}$ state.  This behavior
was not detected on smaller systems. Furthermore, for a large fraction
of the samples, the metastability on $M_\mathrm{s}$ was correlated with
a metastability in the internal energy and in the magnetization
density.  This can be observed, for instance, in the Monte Carlo
history at temperature $T=1.5$ ($H=2.25$) shown in Fig. \ref{fig:sam163} for
a $L=24$ sample. Note that the fluctuations for the internal energy
were huge. However, if one bins 25 consecutive Monte Carlo measurements
(white squares in the central plot) metastability is very clear.

We also learn from Fig. \ref{fig:sam163} that the probability
distribution for the staggered magnetization shows three clear peaks,
one for a disordered state and two for a quasisymmetric
ordered phase. The transition time is of the order of $~10^6$ MC steps
(and tunneling is probably sped up by our use of parallel
tempering). It is then clear that some of the samples may not have had
enough time during the simulation ($2.4\times 10^7$ MC steps) to
perform enough transitions between metastable states to give a correct
value for the mean staggered magnetization, and this explains the
noise we found in observables involving connected functions
(susceptibilities, correlation length and specific heat).

One can estimate the latent heat and the mean energy from Fig.
\ref{fig:sam163} (recall $L=24$): $Q\simeq 0.005$ and $E_+ \simeq E_{-} =E
\simeq 1.36$.  We can introduce these values in the equation from $g_1$ [see
Eq. (\ref{eq:scal_I_ge})]. For small latent heat (we write only the dominant
term) it is possible to obtain \be 1-g=g_1\simeq \frac{Q}{4 E^3} \ee obtaining
$g_1=5 \times 10 ^{-4}$, only at two standard deviations of the $g_1$ value
computed by extrapolating the Binder cumulant.

\section{CONCLUSIONS}\label{CONCLUSIONS}

We have studied the three dimensional diluted antiferromagnetic Ising
model in a magnetic field using equilibrium numerical techniques and
analysis methods.

We have found that the data can be described in the framework of a
second phase transition, and obtained critical exponents are compatible
with those obtained by Ogielski and Huse.~\cite{ogielski} However, the
critical exponent for the order parameter is {\em very} small, which
points to a first-order transition. Note, however, that similarly
small values of this critical exponent were found in ground-state
investigations both for the DAFF~\cite{hartmann1} (at different
dilutions) and for the Gaussian RFIM~\cite{hartmann1,middleton} (these
authors claimed that the
phase transition was continous). Nevertheless, by studying the Binder
cumulant of the energy, we obtained clear indications on our largest
lattices of a weak first-order phase transition.  Furthermore, on our
largest systems, a large number of samples show flip-flops between the
ordered (quasidegenerated) and disordered phases both in the energy
as well as in the order parameter, which again is a strong evidence
for the weakly first-order scenario.

We remark that a complete theory of scaling in disordered first-order
phase transitions (in line with that of
Ref. [\onlinecite{binlau_II}] for ordered systems) is still lacking.
However, we have proposed a set of {\em effective exponents} and
have shown that this scaling accounts for our data.

\acknowledgments

This work has been partially supported by MEC (contracts
Nos. BFM2003-08532, FISES2004-01399, FIS2004-05073, and FIS2006-08533),
by the European Commission (contract No. HPRN-CT-2002-00307), and by
UCM-BCSH. We are grateful to T. J\"{o}rg,
and L. A. Fern\'andez for interesting discussions.
\\
\appendix

\section{Scaling in first-order phase transitions in presence of disorder}\label{APENDICE}
It is straightforward to use the Cauchy-Schwartz inequality to obtain a bound on the
$p$ derivative of an arbitrary observable. This bound will hold
both for first- and second-order phase transitions. Following the lines
of reasoning of 
Ref. [\onlinecite{chayes}] we get
\begin{equation}
\frac{d \overline{\langle A \rangle}}{d p} \le a
\sqrt{\overline{\langle A^2 \rangle}} L^{d/2} \,.
\label{eq:chayes1}
\end{equation}
We recall that $p$ is the dilution of the model, and
$a^2=1/[p(1-p)]$. Notice that this inequality holds for any temperature,
dilution and lattice size.

Assuming now that $\sqrt{\overline{\langle A^2 \rangle}}$ is of the
same order of magnitude of $\overline{\langle A \rangle}$ (which is certainly
the case for the internal energy), we translate Eq. (\ref{eq:chayes1}) into
a bound for the logarithmic derivative:
\begin{equation}
\frac{d \log \overline{\langle A \rangle}}{d p} \le L^{d/2} \,,
\label{eq:chayes2}
\end{equation}
Now, the logarithmic derivative tells us about the width of the critical
region on a finite system. For instance, at a given temperature and magnetic
field, let $p(L)$ be the spin dilution at a susceptibility peak and $p_c$
the thermodynamic limit of any such quantity. We then expect 
$p(L)-p_c\propto L^{-d/2}$.  Notice that the notion of a critical
region permits us to define an effective $\nu$ exponent as $p(L)-p_c
\propto L^{-1/\nu}$. Hence,
\begin{equation}
\nu \ge \frac{2}{d} \,.
\end{equation}
If the coexistence line has finite slope in the ($T$,$p$) plane, it is
clear that the critical width in dilution is proportional to the
critical width in temperature. A similar argument holds for the
derivative with respect to the magnetic field. Thus, the logarithmic
derivative with respect to temperature or magnetic field of $A$ may
diverge (at most) as fast as $L^{d/2}$. So, we have found an upper
bound for the divergences of the specific heat and connected
susceptibilities (both are derivatives of the energy and magnetization
respectively): they cannot diverge, with the lattice size, with an
exponent greater than $d/2$.

Let us also remark that in footnote 7 of Ref. [\onlinecite{chayes}] it
is reported that $\nu=2/d$ for first-order transitions in the presence of
disorder but without an explanation of this fact.

Finally we will show that the Schwartz-Soffer~\cite{schwartz}
inequality also holds in a first-order phase transition scenario. Schwartz
and Soffer show that
\begin{equation}
\hat{\chi}_\mathrm{c}^{M_s}(q) \le \frac{1}{h}
\sqrt{\hat{\chi}_\mathrm{dis}^{M_s}(q)} \,,
\end{equation}
where $q$ is the momentum, $h$ is the standard deviation of the
magnetic field and $\hat{\chi}_\mathrm{c}^{M_s}(q)$ and
$\hat{\chi}_\mathrm{dis}^{M_s}(q)$ are the Fourier transforms of the
connected susceptibility and the disconnected part of it
respectively.~\cite{schwartz}  In order to obtain the inequality, we
introduce the minimum momenta available which is of order $1/L$, and use
$\chi(L)=\hat{\chi}_\mathrm{c}^{M_s}(q_\mathrm{min})$ and
$\chi_\mathrm{dis}(L)=\hat{\chi}_\mathrm{dis}^{M_s}(q_\mathrm{min})$. On the
other hand, in a first-order phase transition, we define the effective
exponents $\eta$ and $\overline{\eta}$ by means of the scaling of the
susceptibilities at the critical point: $\chi(L) \simeq L^{2-\eta}$
and $\chi_\mathrm{dis}(L)\simeq L^{2-\overline{\eta}}$.  All together,
we obtain the Schwartz-Soffer inequality:
\begin{equation}
\frac{2 -\overline{\eta}}{2} \ge 2-\eta \,.
\end{equation}
Note that we have not used the criticality properties of the propagators.


\begin{thebibliography}{99}
\bibitem{fn1}
A few
samples in a $L=64$ lattice were simulated.
\bibitem{fn2}
We recall that they obtained
their estimates of the critical temperature and exponents by
simulating in the paramagnetic phase, without perform finite size
scaling, and by fitting the largest lattice simulated to the standard
law, for example for the susceptibility: $\chi=A (T-T_c) ^{-\gamma}$.
\end{thebibliography}
\end{document}